\documentclass[10pt,conference]{IEEEtran}
\IEEEoverridecommandlockouts
\newcommand{\IEEEAcceptedNotice}{%
To appear in the Proceedings of IEEE CASCON 2026.\\
\copyright~2026 IEEE. Personal use of this material is permitted.  Permission from IEEE must be obtained for all other uses, in any current or future media, including reprinting/republishing this material for advertising or promotional purposes, creating new collective works, for resale or redistribution to servers or lists, or reuse of any copyrighted component of this work in other works.
}

\makeatletter
\def\ps@IEEEtitlepagestyle{%
  \def\@oddfoot{%
    \parbox[b]{0.96\textwidth}{%
      \centering
      \scriptsize
      \IEEEAcceptedNotice
    }%
  }%
  \def\@evenfoot{}%
}
\makeatother

\usepackage{cite}
\usepackage{amsmath,amssymb,amsfonts}
\usepackage{algorithmic}
\usepackage{graphicx}
\usepackage{textcomp}
\usepackage{xcolor}
\usepackage{balance}
\usepackage{url}
\def\BibTeX{{\rm B\kern-.05em{\sc i\kern-.025em b}\kern-.08em
    T\kern-.1667em\lower.7ex\hbox{E}\kern-.125emX}}
\begin{document}

\title{Robustness of Anomaly Detection Models for Industrial Control Systems under Training-Time Data Contamination}

% \author{
% \IEEEauthorblockN{Mustafa Umut Ozbek, Taiwo Ojo, Pooria Madani, Khalil El-Khatib, and Li Yang}
% \IEEEauthorblockA{\textit{Faculty of Business and IT}\\
% \textit{Ontario Tech University}\\
% Oshawa, Ontario, Canada}
% }

\author{\IEEEauthorblockN{
Mustafa Umut Ozbek, Taiwo Ojo, Pooria Madani, Khalil El-Khatib, Li Yang
}
\IEEEauthorblockA{
Ontario Tech University, Oshawa, Ontario, Canada \\
Emails: mustafaumut.ozbek@ontariotechu.net, taiwo.ojo2@ontariotechu.net, \\
pooria.madani@ontariotechu.ca, khalil.el-khatib@ontariotechu.ca, li.yang@ontariotechu.ca
}
}

\maketitle

\begin{abstract}
Machine-learning-based anomaly detection is increasingly used in industrial control systems (ICS), yet most studies assume that detector training data is trustworthy. In practice, training data may be corrupted through compromised logs, labeling errors, manipulated historian records, or unsafe retraining processes. This paper evaluates the robustness of offline ICS anomaly-detection pipelines on the Secure Water Treatment (SWaT) benchmark under training-time contamination. We assess 11 heterogeneous anomaly detectors under three contamination strategies: random injection, similarity-targeted injection, and feature-noise injection. The first two insert attack samples into the nominal training pool, while the third adds bounded Gaussian noise to selected normal training samples. These attacks are contamination-based rather than gradient-driven poisoning methods. Contamination budgets from 1\% to 10\% are evaluated using clean validation and test sets under a unified offline protocol. The results show that robustness is strongly model-dependent and cannot be predicted from clean-data performance alone. Injection-based contamination causes the greatest degradation, particularly for local-density and distance-based detectors, whereas feature-noise contamination has a comparatively limited effect. PCA, SVM, HBOS, and IForest remain relatively stable, while the tuned neural detectors demonstrate intermediate robustness. Overall, the findings highlight the importance of training-data integrity in ML-enabled ICS monitoring, subject to the evaluated dataset, models, and threat assumptions.
\end{abstract}

\begin{IEEEkeywords}
industrial control systems, adversarial machine learning, training-time data contamination, anomaly detection, critical infrastructure
\end{IEEEkeywords}

\section{Introduction}

Industrial control systems (ICS) support critical infrastructure such as water treatment, energy, manufacturing, and digital instrumentation and control systems, where cyber incidents can produce direct operational and safety consequences. As these systems become increasingly connected and data-rich, anomaly-based detection has received growing attention as a complement to signature-based and rule-based monitoring. In particular, anomaly-based detectors are attractive in ICS because normal operating behavior is often easier to collect than comprehensive attack data, and multivariate process measurements can reveal deviations that are difficult to capture with static signatures alone \cite{goh2017rnn,koay2023landscape,ramachandran2021challenges}.

Most existing ICS anomaly-detection studies evaluate models under the implicit assumption that the training data used to build them is trustworthy. In practice, that assumption is often weak. Datasets used for offline detector development may be assembled from historian exports, archived sensor traces, engineering workstations, alarm logs, operator annotations, and automated preprocessing pipelines. Errors, compromise, or manipulation at any stage can contaminate the final training corpus. As a result, a detector that appears strong under clean offline evaluation may still become unreliable if its training data has already been corrupted before deployment \cite{rubinstein2009antidote,cina2022wild}.

This concern is especially important in anomaly-based pipelines trained on normal data only. In such settings, the detector does not learn an explicit supervised boundary between normal and attack classes. Instead, it learns a representation, support region, density structure, geometric profile, or reconstruction pattern of nominal process behavior. Contaminating the normal training pool therefore distorts the learned notion of normality itself. Attack samples injected into nominal training data can weaken the anomaly boundary, while perturbations applied to normal samples can blur the distinction between benign and malicious behavior. Despite this risk, the comparative robustness of anomaly-based ICS detectors under offline training-time contamination remains insufficiently characterized across a broad and diverse set of detector families.

Prior work has demonstrated effective clean-condition detection on the Secure Water Treatment (SWaT) benchmark using recurrent, convolutional, one-class, PCA-based, and reconstruction models \cite{goh2017rnn,kravchik2018cnn,inoue2017unsupervised,kim2020seq2seq,kravchik2022lightweight}. Adversarial ICS studies, however, have concentrated mainly on inference-time evasion \cite{zizzo2020adversarial,10.1145/3427228.3427660,anthi2021adversarial} or poisoning of online-adaptive neural detectors \cite{kravchik2021poisoning,kravchik2022practical}.

We address this gap with a unified benchmark on SWaT \cite{mathur2016swat,goh2017dataset}. Eleven normal-only detectors are tested under three contamination strategies at 1\%, 3\%, 5\%, and 10\% budgets. Thresholds are calibrated on clean validation data and applied unchanged to a clean held-out test set. Our contributions are:
\begin{enumerate}
    \item a comparative offline training-contamination benchmark spanning eleven heterogeneous anomaly detectors;
    \item a common evaluation of random injection, similarity-targeted injection, and feature-noise injection across four budgets;
    \item an analysis that jointly considers F1-score degradation, false-negative rate (FNR), and training cost rather than clean F1-score alone;
    \item a transparent configuration-level evaluation that explicitly reports differences in data representation, tuning depth, training scale, and compute resources.
\end{enumerate}

The remainder of this paper is organized as follows. Section~II reviews related work on ICS anomaly detection, adversarial machine learning, and training-time contamination. Section~III defines the threat model and contamination methodology. Section~IV describes the dataset, detector configurations, and experimental protocol. Section~V presents and discusses the experimental results, and Section~VI concludes the paper and outlines directions for future work.

\section{Related Work}

\subsection{ICS Anomaly Detection}

Anomaly-based detection has become a major line of research in ICS because process behavior is highly structured, while comprehensive attack labels are often limited or costly to obtain. The SWaT testbed and dataset have become widely used benchmarks because they provide a realistic and reproducible platform for evaluating process-aware attack-detection methods \cite{mathur2016swat,goh2017dataset}. Prior SWaT studies include recurrent models trained on normal data \cite{goh2017rnn}, convolutional models \cite{kravchik2018cnn}, unsupervised and one-class approaches \cite{inoue2017unsupervised}, sequence-to-sequence architectures \cite{kim2020seq2seq}, lightweight neural and PCA-based detection \cite{kravchik2022lightweight}, and methodology-oriented work emphasizing preprocessing, thresholding, and evaluation \cite{gomez2020madics}.

Together, these studies support two recurring conclusions. First, anomaly-based monitoring is a natural fit for ICS settings because normal process behavior is repetitive, correlated, and usually easier to collect than diverse attack traces. Second, there is no universally dominant detector family. Broader reviews of machine learning for ICS security reach similar conclusions while identifying persistent challenges in cross-paper comparability, dataset dependence, and deployment realism \cite{koay2023landscape,ramachandran2021challenges}.

Our study also intersects with benchmark design for anomaly and time-series anomaly detection. Prior work shows that detector rankings can shift substantially with metric choice, hyperparameter treatment, and evaluation protocol \cite{campos2016evaluation,schmidl2022comprehensive,wenig2022timeeval}, while time-series studies highlight temporal leakage, split construction, and inconsistent evaluation rules \cite{kim2022rigorous,huet2022local}. These observations reinforce the need to interpret detector performance through both the model family and the evaluation pipeline used to generate the reported results.\footnote{Code and experiment configurations for this paper are publicly available at: \url{https://github.com/ANTS-OntarioTechU/Robustness-Anomaly-Detection-ICS-Data-Contamination}}

\subsection{Adversarial Machine Learning in Industrial Control Systems}

Although adversarial machine learning has received increasing attention in ICS security, most prior work has focused on evasion rather than poisoning. Zizzo \textit{et al.} studied adversarial attacks against time-series intrusion-detection models for ICS and showed that recurrent detectors can be manipulated even when an attacker controls only a subset of variables \cite{zizzo2020adversarial}. Erba \textit{et al.} showed that realistic ICS manipulation must satisfy process and physical constraints, making the attack space substantially different from conventional adversarial examples \cite{10.1145/3427228.3427660}. Anthi \textit{et al.} likewise demonstrated that adversarial perturbations can materially reduce machine-learning-based cyber defenses in industrial environments \cite{anthi2021adversarial}. Collectively, these studies show that strong clean-data detection performance should not be interpreted as evidence of adversarial robustness.

\subsection{Training-Time Poisoning and Contamination of Anomaly Detectors}

Compared with evasion, training-time corruption of ICS anomaly detectors has received much less direct study. Kravchik, Biggio, and Shabtai analyzed poisoning attacks against online-trained neural anomaly detectors for ICS \cite{kravchik2021poisoning}. This work was later extended to a practical evaluation of poisoning attacks on online anomaly detectors, again emphasizing online adaptation as the primary threat surface \cite{kravchik2022practical}. That setting is highly relevant, but differs from the offline training regime used in many operational pipelines, where models are built or periodically refreshed from curated historian logs, archived process traces, and exported datasets before deployment. Online poisoning studies therefore do not fully characterize contamination of the normal training pool during offline model development.

Outside ICS, the broader adversarial machine-learning literature has shown that training-time corruption can substantially degrade anomaly-detection and classification systems. Rubinstein \textit{et al.} demonstrated that poisoning can distort anomaly detectors, particularly PCA-based detectors, by shifting the learned model of normality \cite{rubinstein2009antidote}. Other work developed poisoning attacks against SVMs and deep learning, certified defenses, targeted clean-label attacks, and broader surveys \cite{biggio2012svm,munoz2017backgradient,steinhardt2017certified,shafahi2018poison,cina2022wild}. Recent CPS studies also address provable mitigation and dynamic backdoors \cite{abedzadeh2025mitigating,jiang2026dynamic}. These studies provide the conceptual basis for training-time vulnerability, but primarily address optimization-based poisoning rather than the contamination-style attack families considered here.

\subsection{Research Gap}

Prior ICS studies provide strong evidence that anomaly-based detectors can perform effectively under clean conditions \cite{goh2017rnn,kravchik2018cnn,inoue2017unsupervised,kim2020seq2seq,kravchik2022lightweight,gomez2020madics}, while adversarial ICS studies have focused mainly on evasion or poisoning tied to online retraining \cite{zizzo2020adversarial,10.1145/3427228.3427660,anthi2021adversarial,kravchik2021poisoning,kravchik2022practical}. What remains insufficiently characterized is the comparative robustness of a broad set of anomaly-based ICS detectors under offline training-time contamination within a unified benchmarking framework.

More specifically, to the best of our knowledge, existing literature does not provide a consistent comparison across heterogeneous anomaly-detector families under the same contamination definitions, budgets, threshold-calibration procedure, and clean held-out evaluation protocol. It also remains unclear whether detectors that appear strong on clean benchmark data retain that advantage once the normal training pool is contaminated. The novelty of this study is a unified, configuration-level robustness benchmark across heterogeneous anomaly-detector families under common training pools, contamination budgets, clean calibration data, held-out testing, and evaluation rules.

\section{Threat Model and Contamination Methodology}

\subsection{System and Adversary Model}

We consider an offline anomaly-detection pipeline for ICS under a normal-only training regime. Each detector is trained to model nominal process behavior from a clean training pool, while model selection and threshold calibration use a labeled validation split and final performance is measured on a held-out test split that is never modified during contamination. This setup reflects a practical workflow in which historical process data are curated, preprocessed, and used to develop anomaly detectors before deployment.

The adversary's objective is to degrade the detector's ability to distinguish attacks from normal behavior at test time while preserving the appearance of a plausible training workflow. More specifically, the attacker seeks to increase false negatives, reduce F1-score and recall, and distort the learned notion of normality so that malicious behavior is more likely to be accepted as benign.

We assume a gray-box adversary with training-time access only. The attacker knows the feature representation and general anomaly-detection pipeline, but does not modify the clean validation or test sets and does not interact with the deployed detector at inference time. Depending on the contamination strategy, the adversary either injects attack samples into the nominal training pool or perturbs selected normal training observations. The similarity-targeted strategy assumes more informed sample selection than random contamination, while all evaluated strategies remain weaker than white-box optimization-based attacks. The budget is at most 10\% of the clean training-pool size.

For clarity, ``uncontaminated'' means that no training-time attack is applied to validation or test observations; it does not mean that the labeled validation split is attack-free. The restrictions to clean calibration and test data isolate training-time contamination effects, but may be optimistic relative to workflows in which curation errors or historian compromise propagate beyond the training data.

\begin{figure*}[t]
    \centering
    {\setlength{\unitlength}{1pt}
    \begin{picture}(515,290)
        \put(0,0){\includegraphics[width=515pt,height=290pt]{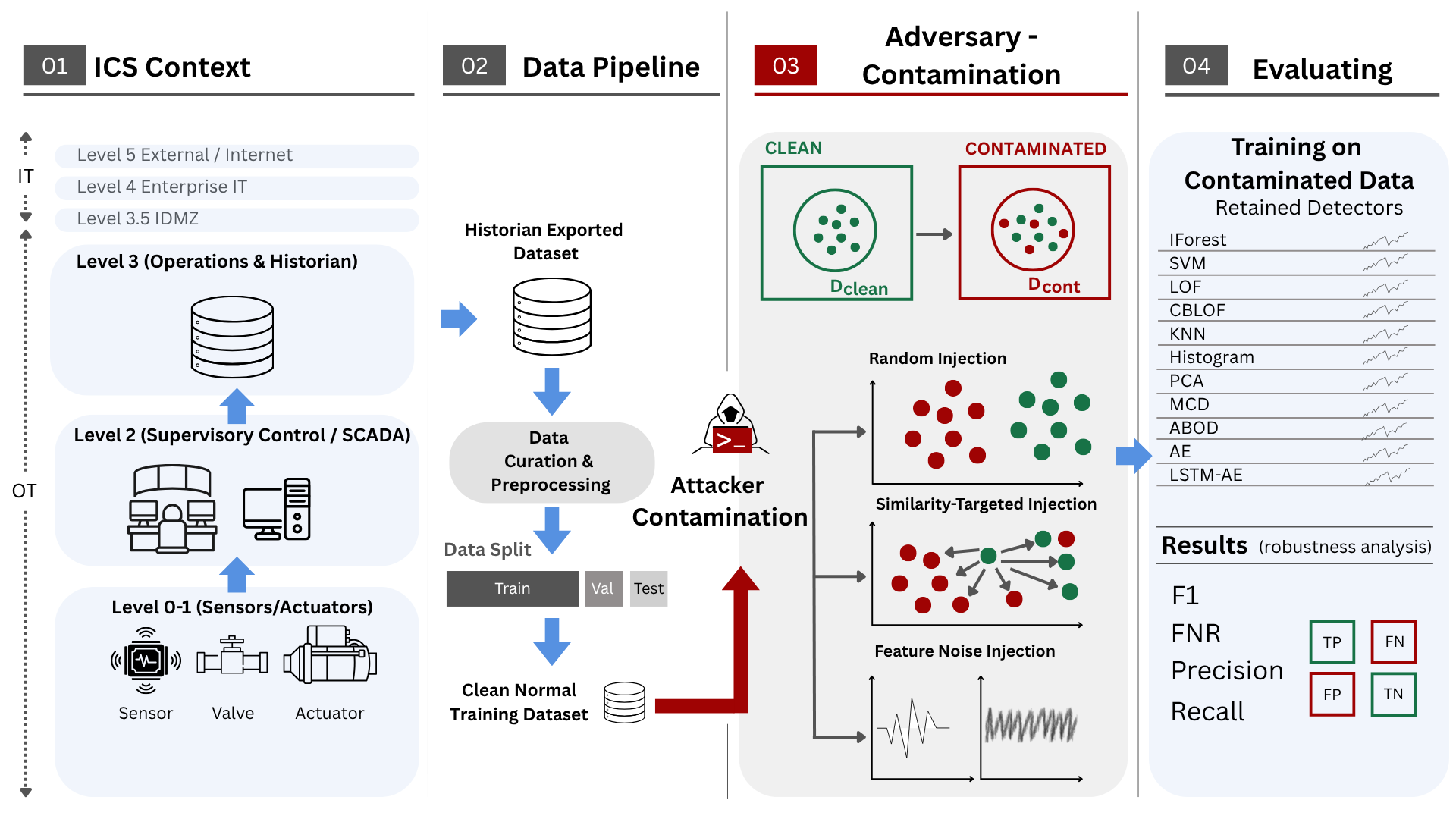}}
    \end{picture}}
    \caption{Offline ICS anomaly-detection workflow. Contamination is confined to the normal training pool; validation and test data remain uncontaminated.}
    \label{fig:pipeline_overview}
\end{figure*}

Offline ICS datasets can absorb attack traces or distorted values through annotation, alignment, curation, export, or preprocessing failures. These failures need not alter the running process; they affect a normal-only detector through data presented as benign. We retain \textit{poisoning} for continuity with the literature, but evaluate contamination-style rather than gradient-driven worst-case attacks. Figure~\ref{fig:pipeline_overview} summarizes the workflow.

Two practical corruption paths motivate the evaluated strategies. First, attack-period observations may be included in a nominal archive through missing labels, event-window misalignment, or unsafe merging of exports. Second, stored normal observations may be modified through historian compromise or preprocessing faults without changing the number of records. Injection attacks model the first path, while feature-noise injection models the second. The benchmark does not claim that these mechanisms exhaust real data-governance failures; they provide repeatable stress tests with a common budget.

\subsection{Contamination Strategies}

Let the clean normal training pool be $D_{\mathrm{clean}}=\{x_i\}_{i=1}^{N}$ and the available attack pool be $A=\{a_j\}_{j=1}^{M}$. For a training-pool-relative budget $p$, the adversary manipulates $k=\lfloor pN\rfloor$ samples. For the injection attacks, appending $k$ samples yields a final injected fraction $k/(N+k)$ (approximately 9.09\% when $p=10\%$); feature noise instead modifies $k$ of the original $N$ samples. Thus, $p$ denotes a common budget relative to the clean pool rather than an identical final-set fraction across attack families.

For the pointwise detectors and AE, the attack pool is the attack-labeled portion of the stratified training fold and is therefore disjoint from their validation and test observations. LSTM-AE uses this pointwise training-fold attack pool to contaminate its separate contiguous normal training block, as detailed in Section~IV-A. This controlled resource isolates training contamination within each implemented protocol but is optimistic relative to end-to-end data-governance failures.

The injected pool should therefore be understood as an evaluation resource rather than an assumption that an operational attacker possesses a perfectly curated attack corpus. It enables controlled comparisons across budgets and seeds. Similarly, the clean validation and test assumption prevents threshold corruption from being conflated with model corruption, but may underestimate failures in pipelines where the same data-quality problem propagates across all splits.

\begin{table}[t]
\caption{Summary of the evaluated training-time contamination strategies.}
\label{tab:attack_summary}
\centering
\scriptsize
\setlength{\tabcolsep}{3pt}
\renewcommand{\arraystretch}{1.08}
\begin{tabular}{|p{1.35cm}|p{3.15cm}|p{1.2cm}|}
\hline
\textbf{Strategy} & \textbf{Training-pool manipulation} & \textbf{Final size} \\
\hline
Random injection & Uniformly sample $k$ observations from attack pool $A$ and append them as nominal & $N+k$ \\
\hline
Similarity-targeted injection & Append the $k$ attack observations closest to the clean centroid & $N+k$ \\
\hline
Feature-noise injection & Perturb $k$ selected clean observations with clipped, scaled Gaussian noise & $N$ \\
\hline
\end{tabular}
\end{table}

\subsubsection{Random Injection}
The attacker uniformly samples $S\subseteq A$ and forms
\begin{equation}
D_{\mathrm{cont}}=D_{\mathrm{clean}}\cup S, \qquad |S|=\min(k,|A|).
\end{equation}
This represents unsafe curation or accidental inclusion of attack traces in data later treated as normal.

Random injection does not exploit detector gradients or model parameters. Its strength comes from relabeling attack behavior implicitly as normal during fitting. It provides a simple baseline for measuring whether a detector's learned support expands around malicious process states.

\subsubsection{Similarity-Targeted Injection}
The clean centroid is $\mu_{\mathrm{clean}}=N^{-1}\sum_i x_i$. Attack samples are ranked by $d_j=\lVert a_j-\mu_{\mathrm{clean}}\rVert_2$, and the $k$ closest samples are inserted. ``Targeted'' therefore denotes centroid-based data selection, not gradient-based optimization or guaranteed worst-case targeting.

This strategy prefers attack observations that appear globally similar to the clean pool after scaling. The centroid is a deliberately model-agnostic selection rule, which allows the same contaminated training set to be evaluated across all detector families. It may be weak for detectors governed by local structure or temporal context, so its results are interpreted as one reproducible similarity heuristic rather than a worst-case targeted attack.

\subsubsection{Feature-Noise Injection}
The attacker selects $Q\subseteq D_{\mathrm{clean}}$, $|Q|=k$, and replaces each selected sample with
\begin{equation}
\tilde{x}_i=\operatorname{clip}_{[0,1]}(x_i+z_i\odot\sigma_{\mathrm{feat}}),
\end{equation}
where $z_i\sim\mathcal{N}(0,0.15^2I)$ and $\sigma_{\mathrm{feat}}$ is the vector of per-feature standard deviations. Noise is added after MinMax scaling. This preserves training-set size while modeling corruption of stored process values.

Clipping maintains the normalized feature range and the per-feature scale factor prevents a common perturbation magnitude from dominating low-variance variables. The experiment fixes the Gaussian magnitude at $0.15$ and varies only the fraction of modified samples. Consequently, a near-null result establishes limited sensitivity to this tested configuration, not general robustness to feature corruption.

The strategies cover unsystematic attack inclusion, centroid-targeted inclusion, and corruption of nominal records. All use the same clean held-out protocol.

\section{Dataset and Experimental Protocol}

\subsection{Dataset and Preprocessing}

SWaT contains multivariate time-series data from normal operation and attack scenarios \cite{mathur2016swat,goh2017dataset}. Timestamp columns are removed, source files are aligned on common features, values are coerced to numeric form, invalid rows are dropped, and seven constant columns (P202, P301, P401, P404, P502, P601, and P603) are removed. After the implemented preprocessing, the pointwise benchmark pool contains 449,919 samples, 44 features, and a 12.14\% attack ratio. MinMax parameters are fitted on training data only and then applied to validation and test data.

Pointwise detectors and AE use a stratified 70/15/15 split of the pooled normal- and attack-file observations. Only normal samples from the training fold are used for detector fitting; labeled validation observations calibrate the F1-oriented threshold. LSTM-AE instead preserves the source-file order: the first 70\% of the normal-operation file is used for training, the next 15\% for early stopping, and the final 15\% supplies normal test observations. The first half of the attack-period file is used for threshold selection and the second half for testing. For contaminated LSTM-AE runs, attack observations are drawn from the attack-labeled portion of the pointwise training fold and appended to the contiguous normal training block. Windows of length $W=30$ are then constructed independently within the training, early-stopping, threshold-selection, and test arrays, so no window crosses a boundary between these arrays. A window is labeled anomalous if any included time step is attacked. This temporal protocol is stricter, but its sequence-level labels are not identical to the pointwise evaluation and must be considered in cross-family comparisons.

Preprocessing is fitted without using validation or test statistics. Timestamp columns are removed, the normal and attack files are aligned on their common process variables, and nonnumeric or invalid rows are discarded before splitting. The seven removed process variables are constant under the available traces and therefore provide no discriminatory variation. For the pointwise protocol, stratification preserves the attack ratio in validation and test data; only the normal portion of the training fold is passed to a detector. For LSTM-AE, preserving order avoids constructing windows from unrelated time points and prevents a window from crossing the boundary between the normal training sequence and the attack period.

The two protocols consequently use different units of evaluation. A pointwise model assigns one score to one 44-dimensional observation. LSTM-AE assigns reconstruction errors to overlapping windows and labels a window anomalous when at least one included time step belongs to an attack interval. We report this distinction because it affects both the number of evaluated instances and the meaning of a false negative. Cross-family comparisons remain useful at the implemented-configuration level, but should not be read as if every detector received an identical sample representation.

\begin{table}[t]
\caption{Pointwise SWaT benchmark pool after preprocessing.}
\label{tab:dataset_summary}
\centering
\footnotesize
\begin{tabular}{|l|r|}
\hline
\textbf{Item} & \textbf{Value} \\
\hline
Processed samples & 449,919 \\
Retained features & 44 \\
Normal / attack samples & 395,298 / 54,621 \\
Attack ratio & 12.14\% \\
Pointwise split & 70\% / 15\% / 15\% \\
Normalization & Train-fitted min-max to $[0,1]$ \\
\hline
\end{tabular}
\end{table}

\subsection{Detectors, Thresholding, and Reproducibility}

The retained models are Isolation Forest (IForest), One-Class SVM, Local Outlier Factor (LOF), Cluster-Based Local Outlier Factor (CBLOF), KNN, Histogram-Based Outlier Score (HBOS), PCA, Minimum Covariance Determinant (MCD), Angle-Based Outlier Detection (ABOD), a feedforward autoencoder (AE), and LSTM-AE \cite{liu2008iforest,scholkopf2001ocsvm,breunig2000lof,he2003cblof,ramaswamy2000knn,goldstein2012hbos,shyu2003pca,rousseeuw1999mcd,kriegel2008abod,sakurada2014autoencoder,kieu2019recurrent}. Classical detectors use PyOD; neural detectors use PyTorch. PCA and SVM are tuned, while other classical models use default or near-default configurations. AE and LSTM-AE use staged hyperparameter searches with multi-seed confirmation.

Table~\ref{tab:detector_summary} summarizes the heterogeneous scoring mechanisms, final retained configurations, and training inputs, allowing contamination sensitivity to be compared across incompatible definitions of normality without duplicating training-scale information.

\begin{table*}[t]
\caption{Detector families, scoring mechanisms, retained configurations, and training inputs used in the comparative benchmark. Classical detectors use PyOD and neural detectors use PyTorch, as described in the text.}
\label{tab:detector_summary}
\centering
\scriptsize
\setlength{\tabcolsep}{2.5pt}
\renewcommand{\arraystretch}{1.08}
\begin{tabular}{|p{2.1cm}|p{3.25cm}|p{7.6cm}|p{2.0cm}|}
\hline
\textbf{Model and family} & \textbf{Scoring mechanism} & \textbf{Retained configuration} & \textbf{Training input} \\
\hline
\textbf{IForest}\newline Tree ensemble & Average path length in random isolation trees & 100 trees; tree subsample 256; contamination 0.05; seed-controlled initialization & Full normal pool \\
\textbf{SVM}\newline Kernel boundary & RBF $\nu$-SVM support boundary, $\nu=0.01$ & RBF one-class SVM; $\nu=0.01$; \texttt{gamma = scale}; contamination 0.05 & 50K subsample \\
\textbf{LOF}\newline Local density & Density ratio relative to a 20-neighbor neighborhood & Novelty mode; 20 neighbors; Minkowski distance; contamination 0.05 & 50K subsample \\
\textbf{CBLOF}\newline Cluster/density & Cluster assignment, cluster size, and distance & Eight clusters; $\alpha=0.9$; $\beta=5$; seed-controlled initialization & Full normal pool \\
\textbf{KNN}\newline Distance & Distance to the fifth-nearest training point & Fifth-nearest-neighbor distance; largest-distance score; Minkowski metric & 50K subsample \\
\textbf{HBOS}\newline Statistical & Feature-wise histogram outlier scores & 10 bins; $\alpha=0.1$; tolerance 0.5; feature-wise histogram scoring & Full normal pool \\
\textbf{PCA}\newline Linear subspace & Reconstruction error at 90\% retained variance & 90\% retained variance; reconstruction-error scoring; seed-controlled initialization & Full normal pool \\
\textbf{MCD}\newline Robust covariance & Robust Mahalanobis distance from a trimmed center & Default support fraction; robust covariance and Mahalanobis-distance scoring & 50K subsample \\
\textbf{ABOD}\newline Geometric & Angle variance with 10 neighbors & Fast variant; 10 neighbors; angle-variance scoring & 50K subsample \\
\textbf{AE}\newline Neural pointwise & Bottleneck reconstruction error & Dimensions $[256,128,64]$; BatchNorm, LeakyReLU, dropout 0.1; Adam $5\times10^{-4}$; batch 2048; MSE; maximum 100 epochs & Full normal pool \\
\textbf{LSTM-AE}\newline Neural sequence & Sequence reconstruction error over $W=30$ & One 256-unit LSTM layer; $W=30$; AdamW $10^{-3}$; batch 256; MSE; maximum 30 epochs & Full sequence set \\
\hline
\end{tabular}
\end{table*}

\subsection{Implementation, Tuning, and Compute Policy}

The classical benchmark uses PyOD implementations with a common normal-only fitting workflow. IForest, HBOS, CBLOF, and PCA use the full pointwise normal training pool, while SVM, LOF, KNN, ABOD, and MCD use a fixed 50,000-point uniform subsample to control super-linear memory or distance-computation costs. IForest uses 100 trees and the default 256-sample tree subsample. LOF uses 20 neighbors, KNN scores distance to the fifth neighbor, CBLOF uses eight clusters, and ABOD uses its fast 10-neighbor configuration. PCA retains 90\% variance, while SVM uses an RBF kernel with $\nu=0.01$ and \texttt{gamma = scale}.

Table~\ref{tab:detector_summary} records the retained settings needed to interpret and reproduce each reported configuration. The PyOD \texttt{contamination} argument is fixed at 0.05 during construction where applicable, but it does not define the final operating threshold. Every detector is converted to binary decisions using the separate validation-calibration procedure described below. Thus, the benchmark's 1--10\% attack budgets describe training-pool manipulation and should not be confused with the library's internal contamination parameter.

Tuning follows a sensitivity-first policy. A classical parameter family is promoted only when a preliminary change improves F1 by at least 0.03 or reduces FNR by at least 0.05 relative to its default slice. Only PCA and SVM cross this promotion threshold; other classical detectors retain default or near-default configurations. This conservative policy avoids tailoring every model extensively to one benchmark, but it also means that the reported ordering is configuration-level evidence rather than a definitive comparison of optimally tuned families.

The neural models receive staged searches because architecture and optimization choices materially affect their performance. AE is evaluated over 426 configurations spanning architecture, optimizer, regularization, threshold, and scaling choices. LSTM-AE is evaluated over 216 configurations spanning sequence length, hidden capacity, and training settings. Final candidates are re-ranked over three seeds using a validation-only composite of clean F1-score and F1-score under 10\% similarity-targeted injection. All hyperparameter screening, staged selection, and multi-seed re-ranking use validation performance only; the held-out test set is accessed only after selecting the final configuration. Table~\ref{tab:tuning_summary} records the resulting policy and configurations.

\begin{table}[t]
\caption{Tuning policy and final model-configuration summary.}
\label{tab:tuning_summary}
\centering
\scriptsize
\setlength{\tabcolsep}{3pt}
\begin{tabular}{|l|p{2.45cm}|p{3.55cm}|}
\hline
\textbf{Group} & \textbf{Search policy} & \textbf{Retained configuration} \\
\hline
Classical & Sensitivity-first; only PCA and SVM promoted & PCA: 90\% variance; SVM: RBF, $\nu=0.01$ \\
\hline
AE & 426 staged configurations; multi-seed confirmation & $[256,128,64]$, LeakyReLU, dropout 0.1, Adam $5\times10^{-4}$, batch 2048 \\
\hline
LSTM-AE & 216 staged configurations; joint sequence/capacity search & $W=30$, hidden size 256, AdamW $10^{-3}$, batch 256 \\
\hline
\end{tabular}
\end{table}

Both neural models minimize mean-squared reconstruction error, use gradient clipping at 1.0, and apply early stopping with patience 15. AE permits up to 100 epochs and LSTM-AE up to 30. The LSTM-AE search also exposes an interaction that a one-factor sweep would miss: $W=30$ and hidden size 256 are not the strongest isolated changes, but their joint configuration ranks highest after staged multi-seed evaluation. Dropout values are not interpreted for the selected single-layer LSTM because PyTorch applies recurrent dropout only between stacked recurrent layers.

Experiments were run on the Narval cluster of the Digital Research Alliance of Canada. Classical jobs used six Intel Xeon CPU cores and 48 GiB of RAM; neural jobs additionally used one NVIDIA A100-SXM4-40 GB GPU. Fixed seeds 42, 123, and 456 are applied to splitting, contamination, initialization, and training. Each run is checkpointed independently before aggregation, supporting restartable execution and consistent generation of the reported tables and figures.

All detectors output continuous anomaly scores. An F1-oriented threshold is selected on validation data and applied unchanged to held-out test data. The primary metric is F1-score; FNR is reported because missed attacks are safety-relevant. Experiments use seeds 42, 123, and 456 and training-pool-relative budgets of 1\%, 3\%, 5\%, and 10\%. Execution produced 430 successful runs: 33 retained clean baselines, 396 retained contamination runs, and one SOD clean baseline used only to document exclusion. SOD was removed from the main benchmark because its clean ROC-AUC indicated inverted score ordering under the evaluated configuration, making contamination-degradation analysis ill-posed \cite{kriegel2009sod}. Its available baseline also showed severe precision collapse (F1 $=0.2013$, precision $=0.1130$, recall $=0.9218$) and a 2011.4-s training time; no SOD contamination result is included in the main analysis. All reported comparative analyses therefore use 429 retained runs.

For every clean or contaminated model within each seed and detector configuration, a new threshold is calibrated using only the same clean labeled validation split. That threshold is then frozen for the corresponding held-out test evaluation. This prevents test-set optimization and measures whether each contaminated model's anomaly-score distribution remains compatible with an operating point selected without contaminated calibration data. Precision, recall, F1-score, FNR, ROC-AUC, and average precision are retained in the experiment outputs; F1 and FNR are emphasized because the former balances the two error directions while the latter directly exposes missed attacks.

\section{Results}

\subsection{Clean Performance and Contamination Robustness}

Table~\ref{tab:main_results} combines clean performance, measured training cost, and worst-case F1 loss. LSTM-AE has the highest clean F1-score ($0.8813$), followed by LOF, ABOD, and AE. Yet the clean-data ranking changes substantially under contamination. LOF, KNN, and ABOD each lose more than 0.64 F1 in their worst condition, while IForest, HBOS, PCA, and SVM remain within 0.068 of their clean baselines. AE and LSTM-AE occupy an intermediate position.

The clean scores are descriptive means over three seeds rather than significance-tested orderings. The leading group is narrow: LSTM-AE, LOF, ABOD, and AE all lie between 0.867 and 0.881. PCA and SVM are mid-table at approximately 0.813--0.814, but require far less training time than ABOD, KNN, AE, or LSTM-AE. Clean evaluation therefore defines the starting point for robustness analysis but does not identify a single operational winner.

The standard deviations also distinguish stable clean estimates from seed-sensitive ones. Most detectors vary by less than 0.01 F1 across the three seeds, whereas LOF and MCD vary more noticeably because their fitting pools are subsampled. These differences are small relative to the largest contamination losses, but they reinforce why degradation is computed from seed-averaged contaminated and clean scores rather than from a single favorable run.

\begin{table*}[t]
\caption{Clean performance (mean $\pm$ SD over three seeds), mean observed training time, and worst-case F1 drop across twelve attack-budget combinations. Worst-case drop is computed from seed-averaged F1 values.}
\label{tab:main_results}
\centering
\scriptsize
\setlength{\tabcolsep}{5pt}
\begin{tabular}{|l|c|c|r|c|}
\hline
\textbf{Detector} & \textbf{Clean F1} & \textbf{Clean FNR} & \textbf{Train (s)} & \textbf{Worst F1 drop} \\
\hline
LSTM-AE & $0.8813 \pm 0.0032$ & $0.2027 \pm 0.0068$ & 139.9 & 0.442 \\
LOF & $0.8761 \pm 0.0208$ & $0.1373 \pm 0.0312$ & 36.5 & 0.660 \\
ABOD & $0.8744 \pm 0.0039$ & $0.2012 \pm 0.0171$ & 249.8 & 0.674 \\
AE & $0.8674 \pm 0.0029$ & $0.2030 \pm 0.0082$ & 117.1 & 0.302 \\
KNN & $0.8482 \pm 0.0016$ & $0.2259 \pm 0.0077$ & 264.2 & 0.648 \\
SVM & $0.8144 \pm 0.0019$ & $0.2922 \pm 0.0015$ & 17.1 & 0.068 \\
PCA & $0.8134 \pm 0.0038$ & $0.3078 \pm 0.0054$ & 1.5 & 0.045 \\
HBOS & $0.7712 \pm 0.0037$ & $0.3425 \pm 0.0088$ & 4.1 & 0.007 \\
CBLOF & $0.7543 \pm 0.0048$ & $0.3874 \pm 0.0034$ & 6.9 & 0.182 \\
IForest & $0.7331 \pm 0.0022$ & $0.4162 \pm 0.0036$ & 4.5 & 0.000 \\
MCD & $0.7260 \pm 0.0129$ & $0.3211 \pm 0.0559$ & 6.2 & 0.126 \\
\hline
\end{tabular}
\end{table*}

\begin{figure}[!b]
    \centering
    \includegraphics[width=\columnwidth]{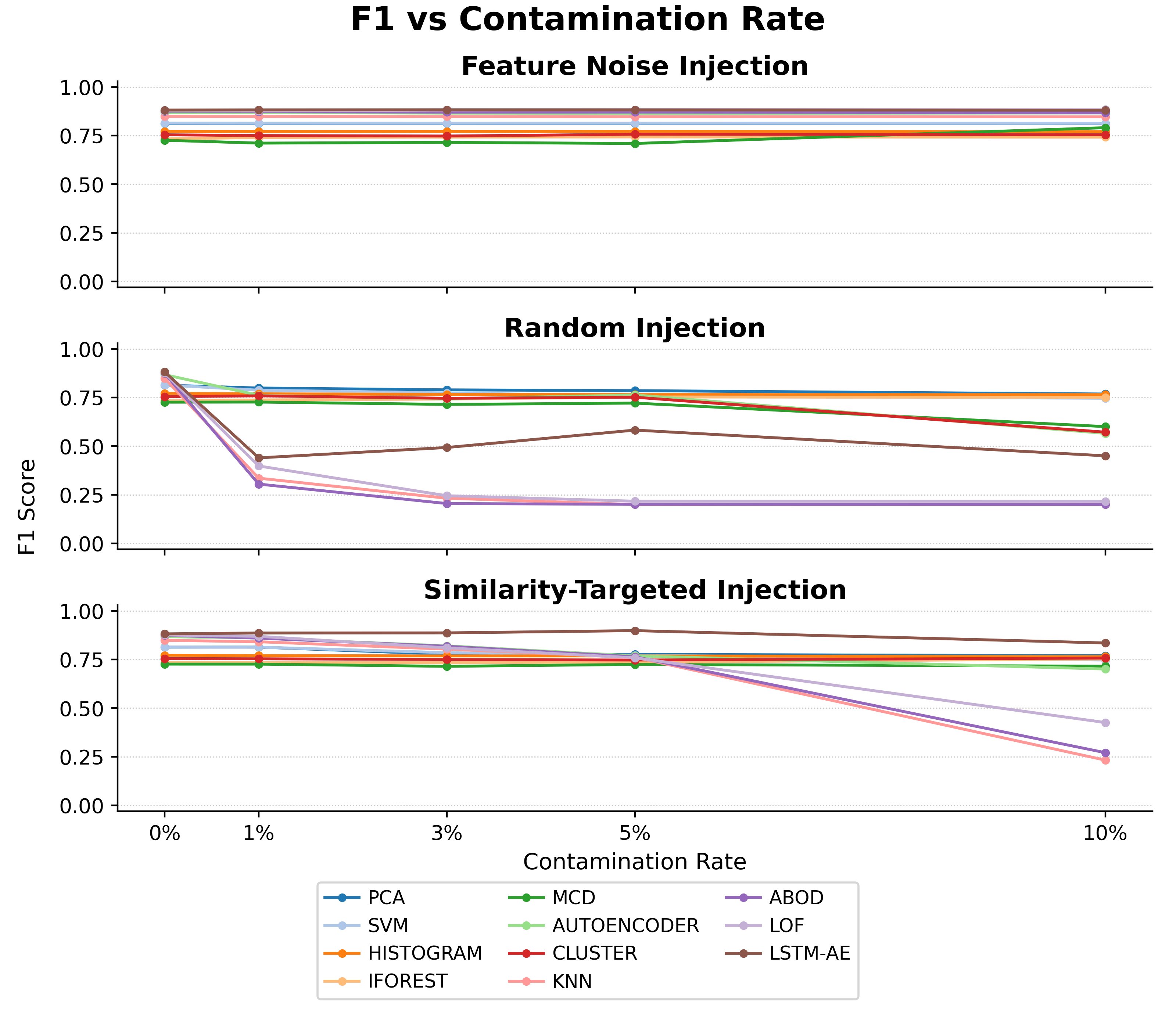}
    \caption{F1-score versus training-pool-relative contamination budget. Injection attacks cause detector-specific degradation; feature noise has a limited effect at the tested magnitude.}
    \label{fig:v4_curves}
\end{figure}

\begin{figure}[!b]
    \centering
    \includegraphics[width=\columnwidth]{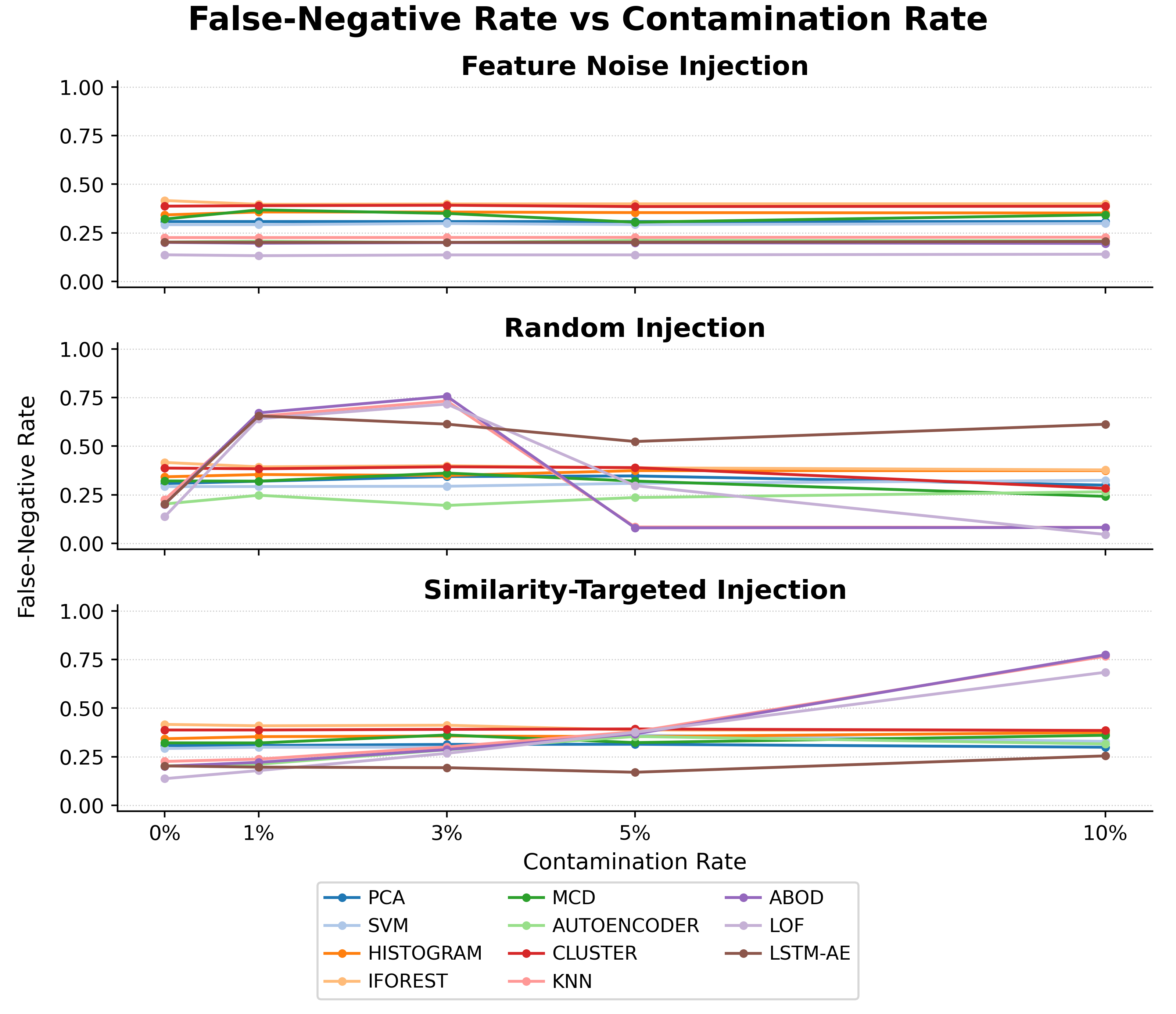}
    \caption{FNR versus training-pool-relative contamination budget. LOF, KNN, ABOD, and LSTM-AE enter the highest missed-detection region under one or more injection settings.}
    \label{fig:v6_fnr}
\end{figure}

Figure~\ref{fig:v4_curves} shows distinct attack responses. Feature noise at $\sigma=0.15$ changes little, whereas 1\% random injection reduces LOF, KNN, ABOD, and LSTM-AE from 0.876, 0.848, 0.874, and 0.881 to approximately 0.398, 0.335, 0.305, and 0.440. AE degrades progressively; similarity-targeted injection becomes severe for selected distance- and density-based detectors at 10\%.

The earliest failures are operationally important because a small clean-pool-relative budget is sufficient to reverse the clean ranking. LOF, KNN, and ABOD begin among the five strongest clean configurations, yet at 1\% random injection each loses more than half of its original F1-score. Their reliance on local neighborhoods, distances, or angles makes them sensitive when attack samples are incorporated into the reference geometry. In contrast, the more aggregated representations of PCA, HBOS, IForest, and SVM change much less under the same evaluated contamination.

The neural response is neither uniformly fragile nor uniformly robust. AE deteriorates gradually and reaches its worst condition at 10\% random injection, where F1 falls to approximately 0.566. LSTM-AE has the highest clean F1-score but drops to approximately 0.440 at only 1\% random injection. Its response to similarity-targeted injection is much milder through 5\%, indicating that the centroid-based pointwise selection rule does not transfer directly into a worst-case sequence attack. The neural results therefore support an intermediate robustness interpretation and demonstrate why attack-family labels should not be treated as equivalent across data representations.

\begin{figure*}[!t]
    \centering
    \includegraphics[width=15cm]{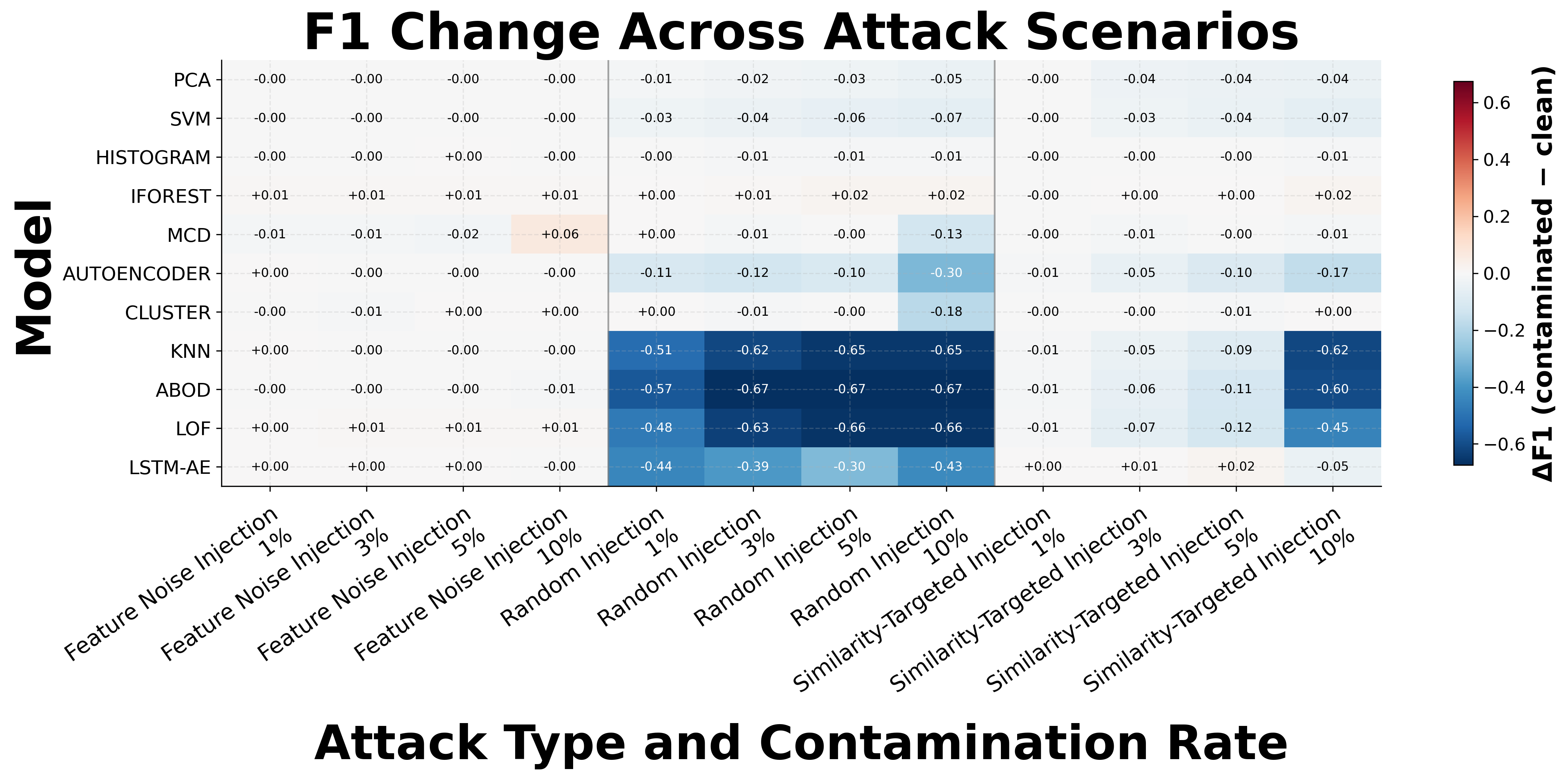}
    \caption{Change in F1-score relative to the clean baseline. Stable classical models remain near zero, while injection attacks expose severe fragility in several high-clean-F1 detectors.}
    \label{fig:v5_heatmap}
\end{figure*}

Figure~\ref{fig:v6_fnr} complements the F1 analysis with missed-detection behavior and is discussed in Section~V-C. In Figs.~\ref{fig:v4_curves}--\ref{fig:v5_heatmap}, the labels HISTOGRAM, CLUSTER, and AUTOENCODER denote HBOS, CBLOF, and AE, respectively; axes labeled ``contamination rate'' report the clean-pool-relative budget $p$ defined in Section~III-B.

The heatmap in Fig.~\ref{fig:v5_heatmap} makes the model dependence explicit. IForest, HBOS, PCA, and SVM remain close to their clean values. Random injection sharply harms LOF, KNN, ABOD, and LSTM-AE across budgets, whereas AE degrades more gradually. Similarity-targeted injection produces delayed failures for LOF, KNN, and ABOD. These results show why a clean-only leaderboard is inadequate for adversarial model selection.

\subsection{Attack-Type Comparison}

The three strategies produce qualitatively different degradation patterns. Random injection is the dominant family overall: it causes the earliest large drops and affects both pointwise neighborhood methods and the sequence model. Because the injected samples span the available attack pool rather than only its center-nearest region, they introduce diverse malicious states into the normal reference data. This pattern is consistent with an expansion or distortion of the learned nominal region for detectors whose scores depend strongly on the training geometry.

Similarity-targeted injection is more detector-dependent. At low budgets it is often less harmful than random injection, but LOF, KNN, and ABOD show delayed failures as the budget increases. The centroid rule selects attack samples that are globally close to the average clean state, yet global proximity does not guarantee local proximity in every detector's representation. The strategy should therefore be viewed as a repeatable stealth-oriented heuristic. It does not establish that random sampling is intrinsically stronger than an optimized targeted attack.

Feature-noise injection remains close to the clean baseline across the grid. Three aspects limit the conclusion: the magnitude is fixed at $\sigma=0.15$, perturbations are clipped to the normalized range, and only selected normal observations are modified. The result shows that the evaluated models tolerate this bounded configuration substantially better than attack-sample injection. It does not show general robustness to biased, coordinated, temporally persistent, or physically constrained sensor corruption.

\subsection{Safety, Cost, and Practical Trade-offs}

Figure~\ref{fig:v6_fnr} shows that FNR should be interpreted together with F1-score rather than in isolation. ABOD, KNN, and LOF exceed approximately 0.70 FNR in at least one contamination setting, while LSTM-AE reaches a lower but still concerning peak under random injection. By contrast, PCA, SVM, HBOS, AE, and MCD remain in the lower-FNR group, and IForest stays comparatively stable despite its lower clean baseline.

At some contamination rates, a lower FNR may reflect overly permissive decision behavior, which reduces missed detections at the cost of precision. For that reason, FNR is most informative when read together with F1-score rather than as a standalone safety measure.

\subsection{Overall Trade-offs Across Performance, Robustness, and Cost}

Taken together, the results show that the ordering by clean-data F1-score is materially different from the ordering implied by robustness and training cost. LSTM-AE, LOF, ABOD, and AE form the strongest clean-data group. However, several of these models remain fragile once the normal training pool is contaminated, particularly LOF, KNN, and ABOD, and to a lesser but still meaningful degree AE and LSTM-AE. By contrast, PCA and SVM do not lead the clean baseline table, yet both combine moderate clean performance with small worst-case F1 loss and much lower training cost. IForest is also notable for its stability across the contamination grid, although its clean baseline remains lower than that of PCA and SVM.

The most defensible practical conclusion is therefore not that one detector dominates on every metric, but that PCA and SVM provide a favorable observed balance among clean performance, contamination robustness, safety-relevant FNR behavior, and computational cost under the evaluated setup. Because hardware and training scales differ, this balance is configuration-specific rather than a formal multi-objective ranking. AE and LSTM-AE remain intermediate rather than robust, and their contamination-time behavior is less dependable than that of the most stable classical configurations.

\subsection{Limitations}

Several limitations should be kept in mind when interpreting the benchmark. First, the study is conducted on a single ICS dataset, so the findings should be interpreted at the benchmark level rather than as population-wide claims about all safety-critical ICS environments. Second, the evaluated attack families are contamination-style training-time scenarios rather than optimization-based worst-case poisoning attacks. Third, the benchmark assumes clean validation and test splits, which isolates training-time contamination effects but may be optimistic relative to workflows in which curation errors propagate beyond the training data. Fourth, the benchmark does not apply equally deep hyperparameter optimization to every detector, so cross-detector comparisons should be interpreted as configuration-level robustness evidence rather than definitive family-level rankings. Finally, the pointwise detectors and sequence model use different split structures: the pointwise random split can place temporally adjacent observations in different folds, whereas LSTM-AE preserves a contiguous protocol. The limited feature-noise effect applies only at $\sigma=0.15$.

\section{Conclusion}

This paper compared eleven anomaly detectors under offline training-time contamination on SWaT. Across three attack families and four budgets, strong clean performance did not imply robustness. LOF, KNN, and ABOD ranked near the top on clean data but degraded most severely after attack-sample injection. AE and LSTM-AE showed intermediate robustness, with random injection as the dominant neural failure mode. PCA, SVM, HBOS, and IForest were comparatively stable; PCA and SVM balanced clean performance, robustness, and cost under the evaluated protocols. Feature noise had limited impact at the tested magnitude, whereas injection was more consequential. These findings motivate integrity controls for offline ICS training data and model selection procedures that explicitly account for contamination. Future work should evaluate additional ICS environments, stronger adaptive attacks, and defenses that jointly optimize accuracy, robustness, and operational cost.

% \IEEEtriggeratref{38}
\balance
\bibliographystyle{IEEEtran}
\bibliography{CASCON_references}

@inproceedings{mathur2016swat,
  author    = {Aditya P. Mathur and Nils Ole Tippenhauer},
  title     = {{SWaT}: A Water Treatment Testbed for Research and Training on ICS Security},
  booktitle = {Proceedings of the 2016 International Workshop on Cyber-physical Systems for Smart Water Networks (CySWater)},
  year      = {2016},
  pages     = {31--36},
  doi       = {10.1109/CySWater.2016.7469060}
}

@incollection{goh2017dataset,
  author    = {Jonathan Goh and Sridhar Adepu and K. N. Junejo and Aditya Mathur},
  title     = {A Dataset to Support Research in the Design of Secure Water Treatment Systems},
  booktitle = {Critical Information Infrastructures Security},
  series    = {Lecture Notes in Computer Science},
  volume    = {10242},
  publisher = {Springer},
  year      = {2017},
  pages     = {88--99},
  doi       = {10.1007/978-3-319-71368-7_8}
}

@inproceedings{goh2017rnn,
  author    = {Jonathan Goh and Sridhar Adepu and Marcus Tan and Zi Shan Lee},
  title     = {Anomaly Detection in Cyber Physical Systems Using Recurrent Neural Networks},
  booktitle = {2017 IEEE 18th International Symposium on High Assurance Systems Engineering (HASE)},
  year      = {2017},
  pages     = {140--145},
  doi       = {10.1109/HASE.2017.36}
}

@inproceedings{kravchik2018cnn,
  author    = {Moshe Kravchik and Asaf Shabtai},
  title     = {Detecting Cyber Attacks in Industrial Control Systems Using Convolutional Neural Networks},
  booktitle = {Proceedings of the 2018 Workshop on Cyber-Physical Systems Security and PrivaCy (CPS-SPC)},
  year      = {2018},
  pages     = {72--83},
  doi       = {10.1145/3264888.3264896}
}

@inproceedings{inoue2017unsupervised,
  author    = {Jun Inoue and Yoriyuki Yamagata and Yuqi Chen and Christopher M. Poskitt and Jun Sun},
  title     = {Anomaly Detection for a Water Treatment System Using Unsupervised Machine Learning},
  booktitle = {2017 IEEE International Conference on Data Mining Workshops (ICDMW)},
  year      = {2017},
  pages     = {1058--1065},
  doi       = {10.1109/ICDMW.2017.149}
}

@incollection{kim2020seq2seq,
  author    = {Jonguk Kim and Jeong{-}Han Yun and Hyoung Chun Kim},
  title     = {Anomaly Detection for Industrial Control Systems Using Sequence-to-Sequence Neural Networks},
  booktitle = {Computer Security},
  series    = {Lecture Notes in Computer Science},
  volume    = {11980},
  publisher = {Springer},
  year      = {2020},
  pages     = {3--18},
  doi       = {10.1007/978-3-030-42048-2_1}
}

@article{kravchik2022lightweight,
  author  = {Moshe Kravchik and Asaf Shabtai},
  title   = {Efficient Cyber Attack Detection in Industrial Control Systems Using Lightweight Neural Networks and {PCA}},
  journal = {IEEE Transactions on Dependable and Secure Computing},
  volume  = {19},
  number  = {4},
  pages   = {2179--2197},
  year    = {2022},
  doi     = {10.1109/TDSC.2021.3050438}
}

@article{gomez2020madics,
  author  = {{\'A}ngel Luis Perales G{\'o}mez and Luis F. Maim{\'o} and Antonio H. Celdr{\'a}n and F{\'e}lix J. Garc{\'\i}a Clemente},
  title   = {{MADICS}: A Methodology for Anomaly Detection in Industrial Control Systems},
  journal = {Symmetry},
  volume  = {12},
  number  = {10},
  pages   = {1583},
  year    = {2020},
  doi     = {10.3390/sym12101583}
}

@article{koay2023landscape,
  author  = {Abigail M. Y. Koay and Ryan K. L. Ko and Henk Hettema and Kenneth Radke},
  title   = {Machine Learning in Industrial Control System Security: Current Landscape, Opportunities and Challenges},
  journal = {Journal of Intelligent Information Systems},
  volume  = {60},
  pages   = {377--405},
  year    = {2023},
  doi     = {10.1007/s10844-022-00753-1}
}

@article{ramachandran2021challenges,
  author  = {Gauthama Raman M. R. and Chuadhry Mujeeb Ahmed and Aditya Mathur},
  title   = {Machine Learning for Intrusion Detection in Industrial Control Systems: Challenges and Lessons from Experimental Evaluation},
  journal = {Cybersecurity},
  volume  = {4},
  number  = {1},
  pages   = {27},
  year    = {2021},
  doi     = {10.1186/s42400-021-00095-5}
}

@inproceedings{zizzo2020adversarial,
  author    = {Giulio Zizzo and Chris Hankin and Sergio Maffeis and Kevin Jones},
  title     = {Adversarial Attacks on Time-Series Intrusion Detection for Industrial Control Systems},
  booktitle = {2020 IEEE 19th International Conference on Trust, Security and Privacy in Computing and Communications (TrustCom)},
  year      = {2020},
  pages     = {899--910},
  doi       = {10.1109/TrustCom50675.2020.00124}
}

@inproceedings{10.1145/3427228.3427660,
  author = {Erba, Alessandro and Taormina, Riccardo and Galelli, Stefano and Pogliani, Marcello and Carminati, Michele and Zanero, Stefano and Tippenhauer, Nils Ole},
  title = {Constrained Concealment Attacks against Reconstruction-based Anomaly Detectors in Industrial Control Systems},
  year = {2020},
  isbn = {9781450388580},
  publisher = {Association for Computing Machinery},
  address = {New York, NY, USA},
  url = {https://doi.org/10.1145/3427228.3427660},
  doi = {10.1145/3427228.3427660},
  booktitle = {Proceedings of the 36th Annual Computer Security Applications Conference},
  pages = {480--495},
  numpages = {16},
  location = {Austin, USA},
  series = {ACSAC '20}
}

@article{anthi2021adversarial,
  author  = {Efthimios Anthi and Lowri Williams and Moritz Rhode and Pete Burnap and Anthony Wedgbury},
  title   = {Adversarial Attacks on Machine Learning Cybersecurity Defences in Industrial Control Systems},
  journal = {Journal of Information Security and Applications},
  volume  = {58},
  pages   = {102717},
  year    = {2021},
  doi     = {10.1016/j.jisa.2020.102717}
}

@inproceedings{kravchik2021poisoning,
  author    = {Moshe Kravchik and Battista Biggio and Asaf Shabtai},
  title     = {Poisoning Attacks on Cyber Attack Detectors for Industrial Control Systems},
  booktitle = {Proceedings of the 36th Annual ACM Symposium on Applied Computing (SAC)},
  year      = {2021},
  pages     = {116--125},
  doi       = {10.1145/3412841.3441892}
}

@article{kravchik2022practical,
  author  = {Moshe Kravchik and Luca Demetrio and Battista Biggio and Asaf Shabtai},
  title   = {Practical Evaluation of Poisoning Attacks on Online Anomaly Detectors in Industrial Control Systems},
  journal = {Computers \& Security},
  volume  = {122},
  pages   = {102901},
  year    = {2022},
  doi     = {10.1016/j.cose.2022.102901}
}

@inproceedings{rubinstein2009antidote,
  author    = {Benjamin I. P. Rubinstein and Blaine Nelson and Ling Huang and Anthony D. Joseph and Shing-hon Lau and Satish Rao and Nina Taft and J. D. Tygar},
  title     = {{ANTIDOTE}: Understanding and Defending Against Poisoning of Anomaly Detectors},
  booktitle = {Proceedings of the 9th ACM SIGCOMM Conference on Internet Measurement (IMC)},
  year      = {2009},
  pages     = {1--14},
  doi       = {10.1145/1644893.1644895}
}

@inproceedings{biggio2012svm,
  author    = {Battista Biggio and Blaine Nelson and Pavel Laskov},
  title     = {Poisoning Attacks against Support Vector Machines},
  booktitle = {Proceedings of the 29th International Conference on Machine Learning (ICML)},
  year      = {2012},
  pages     = {1467--1474}
}

@inproceedings{munoz2017backgradient,
  author    = {Luis Mu{\~n}oz-Gonz{\'a}lez and Battista Biggio and Ambra Demontis and Andrea Paudice and Vasin Wongrassamee and Emil C. Lupu and Fabio Roli},
  title     = {Towards Poisoning of Deep Learning Algorithms with Back-gradient Optimization},
  booktitle = {Proceedings of the 10th ACM Workshop on Artificial Intelligence and Security (AISec)},
  year      = {2017},
  pages     = {27--38},
  doi       = {10.1145/3128572.3140451}
}

@inproceedings{liu2008iforest,
  author    = {Fei Tony Liu and Kai Ming Ting and Zhi-Hua Zhou},
  title     = {Isolation Forest},
  booktitle = {2008 Eighth IEEE International Conference on Data Mining},
  year      = {2008},
  pages     = {413--422},
  doi       = {10.1109/ICDM.2008.17}
}

@article{scholkopf2001ocsvm,
  author  = {Bernhard Sch{\"o}lkopf and John C. Platt and John Shawe-Taylor and Alex J. Smola and Robert C. Williamson},
  title   = {Estimating the Support of a High-Dimensional Distribution},
  journal = {Neural Computation},
  volume  = {13},
  number  = {7},
  pages   = {1443--1471},
  year    = {2001},
  doi     = {10.1162/089976601750264965}
}

@inproceedings{breunig2000lof,
  author    = {Markus M. Breunig and Hans-Peter Kriegel and Raymond T. Ng and J{\"o}rg Sander},
  title     = {{LOF}: Identifying Density-Based Local Outliers},
  booktitle = {Proceedings of the 2000 ACM SIGMOD International Conference on Management of Data},
  year      = {2000},
  pages     = {93--104},
  doi       = {10.1145/342009.335388}
}

@article{he2003cblof,
  author  = {Zengyou He and Xiaofei Xu and Shengchun Deng},
  title   = {Discovering Cluster-Based Local Outliers},
  journal = {Pattern Recognition Letters},
  volume  = {24},
  number  = {9--10},
  pages   = {1641--1650},
  year    = {2003},
  doi     = {10.1016/S0167-8655(03)00003-5}
}

@inproceedings{ramaswamy2000knn,
  author    = {Sridhar Ramaswamy and Rajeev Rastogi and Kyuseok Shim},
  title     = {Efficient Algorithms for Mining Outliers from Large Data Sets},
  booktitle = {Proceedings of the 2000 ACM SIGMOD International Conference on Management of Data},
  year      = {2000},
  pages     = {427--438},
  doi       = {10.1145/342009.335437}
}

@inproceedings{goldstein2012hbos,
  author    = {Markus Goldstein and Andreas Dengel},
  title     = {Histogram-Based Outlier Score ({HBOS}): A Fast Unsupervised Anomaly Detection Algorithm},
  booktitle = {KI-2012: Poster and Demo Track},
  year      = {2012}
}

@inproceedings{shyu2003pca,
  author    = {Meng-Lin Shyu and Shu-Ching Chen and Kanoksri Sarinnapakorn and Liwu Chang},
  title     = {A Novel Anomaly Detection Scheme Based on Principal Component Classifier},
  booktitle = {Proceedings of the IEEE Foundations and New Directions of Data Mining Workshop},
  year      = {2003},
  pages     = {172--179}
}

@article{rousseeuw1999mcd,
  author  = {Peter J. Rousseeuw and Katrien Van Driessen},
  title   = {A Fast Algorithm for the Minimum Covariance Determinant Estimator},
  journal = {Technometrics},
  volume  = {41},
  number  = {3},
  pages   = {212--223},
  year    = {1999},
  doi     = {10.1080/00401706.1999.10485670}
}

@inproceedings{kriegel2008abod,
  author    = {Hans-Peter Kriegel and Matthias Schubert and Arthur Zimek},
  title     = {Angle-Based Outlier Detection in High-dimensional Data},
  booktitle = {Proceedings of the 14th ACM SIGKDD International Conference on Knowledge Discovery and Data Mining},
  year      = {2008},
  pages     = {444--452},
  doi       = {10.1145/1401890.1401946}
}

@incollection{kriegel2009sod,
  author    = {Hans-Peter Kriegel and Peer Kr{\"o}ger and Erich Schubert and Arthur Zimek},
  title     = {Outlier Detection in Axis-Parallel Subspaces of High Dimensional Data},
  booktitle = {Advances in Knowledge Discovery and Data Mining},
  series    = {Lecture Notes in Computer Science},
  volume    = {5476},
  publisher = {Springer},
  year      = {2009},
  pages     = {831--838},
  doi       = {10.1007/978-3-642-01307-2_86}
}

@inproceedings{sakurada2014autoencoder,
  author    = {Mayu Sakurada and Takehisa Yairi},
  title     = {Anomaly Detection Using Autoencoders with Nonlinear Dimensionality Reduction},
  booktitle = {Proceedings of the MLSDA 2014 2nd Workshop on Machine Learning for Sensory Data Analysis},
  year      = {2014},
  pages     = {4:1--4:11},
  doi       = {10.1145/2689746.2689747}
}

@inproceedings{kieu2019recurrent,
  author    = {Tung Kieu and Bin Yang and Chenjuan Guo and Christian S. Jensen},
  title     = {Outlier Detection for Time Series with Recurrent Autoencoder Ensembles},
  booktitle = {Proceedings of the Twenty-Eighth International Joint Conference on Artificial Intelligence},
  year      = {2019},
  pages     = {2725--2732},
  doi       = {10.24963/ijcai.2019/378}
}

@article{campos2016evaluation,
  title={On the Evaluation of Unsupervised Outlier Detection: Measures, Datasets, and an Empirical Study},
  author={Campos, Guilherme Oliveira and Zimek, Arthur and Sander, J{\"o}rg and Campello, Ricardo J. G. B. and Micenkov{\'a}, Barbora and Schubert, Erich and Assent, Ira and Houle, Michael E.},
  journal={Data Mining and Knowledge Discovery},
  volume={30},
  number={4},
  pages={891--927},
  year={2016},
  doi={10.1007/s10618-015-0444-8}
}

@article{schmidl2022comprehensive,
  title={Anomaly Detection in Time Series: A Comprehensive Evaluation},
  author={Schmidl, Sebastian and Wenig, Phillip and Papenbrock, Thorsten},
  journal={Proceedings of the VLDB Endowment},
  volume={15},
  number={9},
  pages={1779--1797},
  year={2022},
  doi={10.14778/3538598.3538602}
}

@article{wenig2022timeeval,
  title={TimeEval: A Benchmarking Toolkit for Time Series Anomaly Detection Algorithms},
  author={Wenig, Phillip and Schmidl, Sebastian and Papenbrock, Thorsten},
  journal={Proceedings of the VLDB Endowment},
  volume={15},
  number={12},
  pages={3678--3681},
  year={2022},
  doi={10.14778/3554821.3554873}
}

@inproceedings{kim2022rigorous,
  title={Towards a Rigorous Evaluation of Time-Series Anomaly Detection},
  author={Kim, Siwon and Choi, Kukjin and Choi, Hyun-Soo and Lee, Byunghan and Yoon, Sungroh},
  booktitle={Proceedings of the AAAI Conference on Artificial Intelligence},
  volume={36},
  number={7},
  pages={7194--7201},
  year={2022},
  doi={10.1609/aaai.v36i7.20680}
}

@inproceedings{huet2022local,
  title={Local Evaluation of Time Series Anomaly Detection Algorithms},
  author={Huet, Alexis and Navarro, Jose Manuel and Rossi, Dario},
  booktitle={Proceedings of the 28th ACM SIGKDD Conference on Knowledge Discovery and Data Mining},
  pages={635--645},
  year={2022},
  doi={10.1145/3534678.3539339}
}

@inproceedings{steinhardt2017certified,
  title={Certified Defenses for Data Poisoning Attacks},
  author={Steinhardt, Jacob and Koh, Pang Wei and Liang, Percy},
  booktitle={Advances in Neural Information Processing Systems},
  year={2017}
}

@inproceedings{shafahi2018poison,
  title={Poison Frogs! Targeted Clean-Label Poisoning Attacks on Neural Networks},
  author={Shafahi, Ali and Huang, W. Ronny and Najibi, Mahyar and Suciu, Octavian and Studer, Christoph and Dumitras, Tudor and Goldstein, Tom},
  booktitle={Advances in Neural Information Processing Systems},
  year={2018}
}

@article{cina2022wild,
  title={Wild Patterns Reloaded: A Survey of Machine Learning Security against Training Data Poisoning},
  author={Cin{\`a}, Antonio Emanuele and Grosse, Kathrin and Demontis, Ambra and Vascon, Sebastiano and Zellinger, Werner and Moser, Bernhard A. and Oprea, Alina and Biggio, Battista and Pelillo, Marcello and Roli, Fabio},
  journal={ACM Computing Surveys},
  volume={55},
  number={13s},
  pages={1--39},
  year={2023},
  doi={10.1145/3585385}
}

@article{abedzadeh2025mitigating,
  author  = {Sahar Abedzadeh and Shameek Bhattacharjee},
  title   = {Mitigating Impact of Data Poisoning Attacks on {CPS} Anomaly Detection with Provable Guarantees},
  journal = {Information},
  volume  = {16},
  number  = {6},
  pages   = {428},
  year    = {2025},
  doi     = {10.3390/info16060428}
}

@article{jiang2026dynamic,
  author  = {Qingchao Jiang and Yu Zu and Zhiying Zhu and Weimin Zhong and Yiran Xu and Zhenxing Qian and Xinpeng Zhang},
  title   = {Dynamic Stealthy Backdoor Attack against Anomaly Detectors in Industrial Control Systems},
  journal = {Information Sciences},
  volume  = {735},
  pages   = {123066},
  year    = {2026},
  doi     = {10.1016/j.ins.2025.123066}
}

\end{document}